\documentclass[runningheads]{llncs_template/llncs}

\usepackage{graphicx}
\usepackage{amsmath}
\usepackage{xcolor}
\usepackage{amssymb}
\usepackage{fontawesome}
\usepackage{cite}

\DeclareMathOperator*{\argmin}{arg\,min}

\title{Towards Development with Multi-Version Models: Detecting Merge Conflicts and Checking Well-Formedness\thanks{This work was developed mainly in the course of the project modular and incremental Global Model Management (project number 336677879) funded by the DFG.
\\
This preprint has not undergone peer review (when applicable) or any post-submission improvements or corrections. The Version of Record of this contribution is published in Graph Transformation, 15th International Conference, Proceedings, and is available online at \url{https://doi.org/10.1007/978-3-031-09843-7}.
}}

\author{Matthias Barkowsky\textsuperscript{\faEnvelopeO} \and Holger Giese}

\authorrunning{M. Barkowsky \and H. Giese}

\institute{Hasso-Plattner Institute at the University of Potsdam\\
Prof.-Dr.-Helmert-Str. 2-3, D-14482 Potsdam, Germany\\
\email{\{matthias.barkowsky,holger.giese\}@hpi.de}}

\begin{document}

\newcommand{\mcomment}[1]{\marginpar{\color{blue}#1\color{black}}}
\newcommand{\gspan}[3]{\ensuremath{(#1\leftarrow#2\rightarrow#3)}}

\maketitle

\begin{abstract}
Developing complex software requires that multiple views and versions of the software can be developed in parallel and merged as supported by views and managed by version control systems.
In this context, this paper considers monitoring merging and related consistency problems permanently at the level of models and abstract syntax to permit early and frequent conflict detection while developing in parallel. The presented approach introduces multi-version models based on typed graphs that permit to store changes and multiple versions in one graph in a compact form and allow (1) to study well-formedness for all versions without the need to extract each version individually, (2) to report all possible merge conflicts without the need to merge all pairs of versions, and (3) to report all violations of well-formedness conditions that will result for merges of any two versions independent of any merge decisions without the need to merge all pairs of versions. 
The paper defines the related concepts and algorithms operating on multi-version models, proves their correctness w.r.t.~the usually employed three-way-merge, and reports on preliminary experiments concerning the scalability. 
\end{abstract}


\section{Introduction}

Developing complex software nowadays requires that multiple views and versions of the software can be developed in parallel and merged as supported by views and managed by version control systems \cite{Fruehauf&Zeller1999}. For complex software, living with inconsistencies at least temporarily is inevitable, as enforcing consistency may lead to loss of important information \cite{Finkelstein+1994} and is hence neither always possible nor desirable. However, working with multiple versions in parallel and changing each version on its own for longer periods of time can introduce substantial conflicts that are difficult and expensive to resolve. Therefore, it is necessary to manage consistency when combining views and versions using merge approaches \cite{Fruehauf&Zeller1999,Mens2005}.

In this context, this paper considers monitoring merging and related consistency problems permanently at the level of models and abstract syntax. This aims to permit early and frequent conflict detection while developing in parallel, as suggested in approaches to detect conflicts early and to enable collaboration to manage conflicts and their risks \cite{Brun+2013}.

The presented approach therefore introduces multi-version models based on typed graphs, which permit to store changes and multiple versions in one graph in a compact form and allow to study the different versions and their merge combinations. The following capabilities are considered:
(1) Study well-formedness for all versions at once without the need to extract and explicitly consider each version individually.
(2) Report all possible merge conflicts that may result for merges of any two versions without the need to extract and explicitly merge all pairs of versions.  
(3) Report all violations of well-formedness conditions that will result for merges of any two versions independent of any merge decisions without the need to extract and explicitly merge all pairs of versions.

The approach thus promises to support early conflict detection and collaboration for managing conflicts and their risks, while not having to decide how to later merge conflicting versions. The technique also aims for a better scalability in case there are many versions that are considered in parallel.

Furthermore, the developed multi-version models permit to study the phenomena of versions, merging, and well-formedness conditions in the unifying framework of typed graphs. This enables us to 
(a) formulate algorithms that can obtain several analysis results without the need to consider a specific version, merge of a pair of versions, or strategy for conflict resolution and
(b) prove that the algorithms compute the same results as if we would explicitly consider all specific versions, merges of pairs of versions, or strategies for conflict resolution.

The paper defines the related concepts and algorithms operating on multi-version models, proves their correctness w.r.t.~the usually employed three-way-merge, and reports on first experiments concerning the scalability. 
In Section 2, we summarize the preliminaries of the presented approach, including basic definitions for typed graphs, well-formedness conditions, and graph modifications.
Then, as a baseline, single-version models in the form of typed graphs with well-formedness conditions are defined in Section 3, before multi-version models are introduced in Section 4. 
Determining all merge conflicts and checking well-formedness for all merge results based on multi-version models is then considered in Section 5.
Results of first experiments for our prototypical implementation of the algorithms are presented in Section 6. 
Finally, the conclusions of the paper and an outlook of planned future work are presented in Section 7.
For the submitted review version, more detailed proofs and the conditions used in the evaluation are in addition presented in an appendix.


\section{Preliminaries} \label{seq:prerequisites}

We briefly reiterate the basic concepts of graphs, graph modifications, and well-formedness conditions used in the remainder of the paper.

A graph $G = (V^G, E^G, s^G, t^G)$ consists of a set of nodes $V^G$, a set of edges $E^G$ and two functions $s^G: E^G \rightarrow V^G$ and $t^G: E^G \rightarrow V^G$ assigning each edge its source and target, respectively. We assume that graph elements have identities and source and target of an edge are invariant if an edge is part of multiple graphs, that is, for two graphs $G$ and $H$ and an edge $e \in E^{G} \cap E^{H}$, it holds that $s^{G}(e) = s^{H}(e)$ and $t^{G}(e) = t^{H}(e)$. This also implies that, in the context of this paper, $(V^{G}= V^{H} \wedge E^{G} = E^{H}) \rightarrow (G = H)$.

A graph morphism $m: G \rightarrow H$ is given by a pair of functions $m^V: V^G \rightarrow V^H$ and $m^E: E^G \rightarrow E^H$ that map elements from $G$ to elements from $H$ such that $s^H \circ m^E = m^E \circ s^G$ and $t^H \circ m^E = m^E \circ t^G$ \cite{Ehrig+2006}.

A graph $G$ can be typed over a type graph $TG$ via a typing morphism $\mathit{type}: G \rightarrow TG$, forming the typed graph $G^T = (G, \mathit{type}^G)$. A typed graph morphism between two typed graphs $G^T = (G, \mathit{type}^G)$ and $H^T = (H, \mathit{type}^H)$ with the same type graph then denotes a graph morphism $m^T: G \rightarrow H$ such that $\mathit{type}^G = \mathit{type}^H \circ m^T$. A (typed) graph morphism is a monomorphism iff its functions are both surjective and injective.

\begin{figure}
\centering
\includegraphics[width=\textwidth]{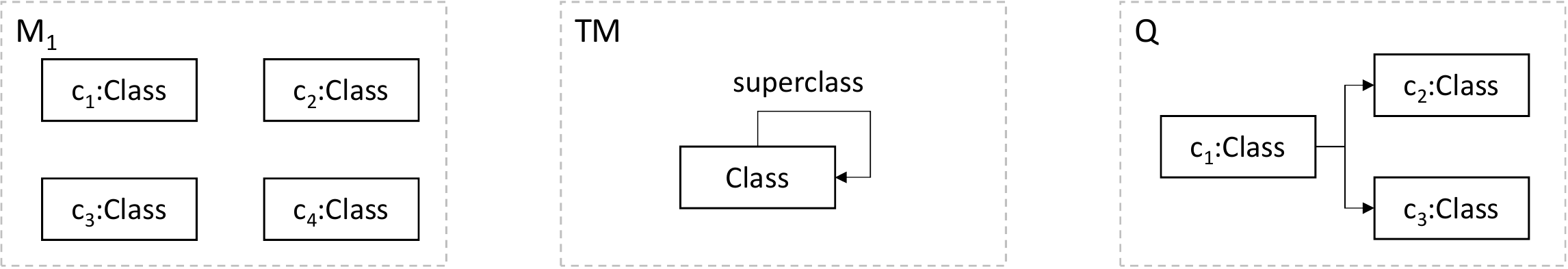}
\caption{Example graph, type graph, and violation pattern} \label{fig:example_graphs}
\end{figure}

Figure \ref{fig:example_graphs} shows an example typed graph $M_1$ and associated type graph $TM$ from the software development domain. $M_1$ represents an abstract syntax graph for a program written in an object-oriented language that contains four classes represented by nodes, with edges representing superclass relationships.

The structure of a typed graph $G$ can be restricted by a well-formedness condition $\phi$, which in the context of this paper is characterized by a typed graph $Q$ typed over the same type graph. $G$ then satisfies the condition $\phi$, denoted $G \models \phi$, iff there exists no monomorphism $m: Q \rightarrow G$. We also call such monomorphisms \emph{matches} and $Q$ the \emph{violation pattern} of $\phi$.

Figure \ref{fig:example_graphs} shows a violation pattern $Q$ for an example well-formedness constraint that forbids a class having two outgoing superclass relationships.

A graph modification as defined by Taentzer et al. \cite{taentzer2014fundamental} formalizes the difference between two graphs $G$ and $H$ and is characterized by an intermediate graph $K$ and a span of monomorphisms \gspan{G}{K}{H}. In this paper, we assume that the two morphisms are always partial identities. The graph $K$ then characterizes the subgraph that is preserved through the modification, whereas all elements in $G$ that are not in $K$ are deleted and elements in $H$ but not in $K$ are created.

\begin{figure}
\centering
\includegraphics[width=\textwidth]{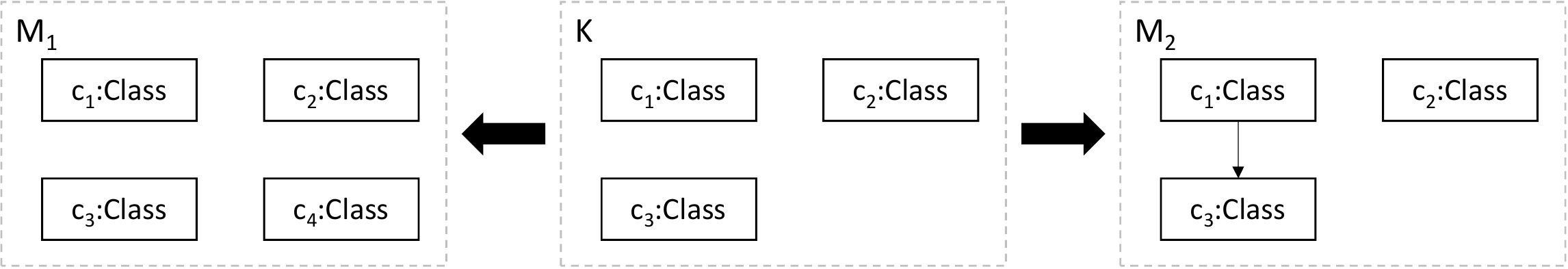}
\caption{Example graph modification} \label{fig:example_modification}
\end{figure}

Figure \ref{fig:example_modification} shows an example graph modification from the graph $M_1$ from Figure \ref{fig:example_graphs} to a new graph $M_2$, where a superclass edge from class $c_1$ to class $c_3$ is created and the class $c_4$ is deleted. The morphisms are implied by node labels.

Graphs and graph modifications correspond to versions and differences in conventional, line-based version control systems like Git \cite{git}, where versions of a development artifact and intermediate differences form a directed acyclic graph.


\section{Single-Version Models}

In this paper, we consider models in the form of typed graphs that are required to adhere to a set of well-formedness conditions. Effectively, the combination of type graph and well-formedness conditions then acts as a metamodel with potential further constraints. Note that attributes, as usually employed in real-world models, can in this context be modeled as dedicated nodes \cite{Heckel+2002}.

For $\Phi$ the set of well-formedness conditions, a model $M_i$ is \emph{well-formed} iff
$
  \forall \phi \in \Phi: M_i \models \phi
  .
$
We assume $\mathop{pcheck}(M_i,\phi)$ to report all violations to property $\phi$ with violation pattern $Q$ for model $M_i$ in the form of matches for $Q$, essentially realizing $\models$ as $\mathop{pcheck}(M_i,\phi) = \emptyset  \iff M_i \models \phi$. If violations exist, the model $M_i$ is also called \emph{ill-formed}.

For the notion of models as typed graphs, model modifications correspond to graph modifications as presented in Section \ref{seq:prerequisites}. We say a model modification \gspan{M_i}{K}{M_j} with identity morphisms is \emph{maximally preserving} iff it does not delete and recreate identical elements. Formally, $K = (V^{M_i} \cap V^{M_j}, E^{M_i} \cap E^{M_j}, s^K, t^K)$, where $s^K$ and $t^K$ are uniquely defined assuming invariant edge sources and targets. Consequently, for two models $M_i$ and $M_j$, the maximally preserving model modification \gspan{M_i}{K}{M_j} is uniquely defined.

For a set of model modifications $\Delta^{M_{\{1, \dots, n\}}}$ between models $M_{\{1, \dots, n\}} = \{M_1, \dots, M_n\}$, with $\forall \gspan{G}{K}{H} \in \Delta^{M_{\{1, \dots, n\}}}: G \in M_{\{1, \dots, n\}} \wedge H \in M_{\{1, \dots, n\}}$, we can define the set of predecessors $\mathop{pre}(i) \subset M_{\{1, \dots, n\}}$ of a version $M_i$ as the set of versions $M_j$ such that there exists a sequence of model modifications \gspan{M_{x_1}}{K_{x_1}}{M_{x_{2}}}, \gspan{M_{x_{2}}}{K_{x_{2}}}{M_{x_{3}}}, \dots, \gspan{M_{x_{n - 1}}}{K_{x_{n - 1}}}{M_{x_{n}}} where $x_1 = j$, $x_{n} = i$, and $\gspan{M_{x_k}}{K_{x_k}}{M_{x_{k + 1}}} \in \Delta^{M_{\{1, \dots, n\}}}$ for $1 \leq k < n$.

$\Delta^{M_{\{1, \dots, n\}}}$ describes a \emph{correct model versioning} if all morphisms in the individual model modifications are partial identities, all model modifications are maximally preserving, the $\mathop{pre}$ relation is acyclic and there exists a model $M_\alpha$ such that $M_\alpha \in \mathop{pre}(i)$ for all models $M_i \neq M_\alpha$. Effectively, a correct model versioning describes a directed acyclic graph of model versions $M_{\{1, \dots, n\}}$ that are derived from an original model $M_\alpha$ via the model modifications in $\Delta^{M_{\{1, \dots, n\}}}$, and therefore closely corresponds to the versioning of some development artifact in a conventional version control system.

Taentzer et al. \cite{taentzer2014fundamental} define a merge operation for model modifications $m_1 = \gspan{M_c}{K_i}{M_i}$ and $m_2 = \gspan{M_c}{K_j}{M_j}$ with common source $M_c$, which unifies $m_1$ and $m_2$ into a merged model modification $m_m = \mathop{merge}(m_1, m_2) = \gspan{M_c}{K_m}{M_m}$. We denote the merged model by $M_m = \mathop{merge_G}(m_1, m_2)$. This merge operation is similar to a three-way-merge in conventional version control systems\cite{Mens2005}, since $m_m$ in the default case (i) preserves an element $x \in M_c$ iff it is preserved by both $m_1$ and $m_2$ (ii) deletes an element $x \in M_c$ iff it is deleted by $m_1$ or $m_2$ (iii) creates an element $x \in M_m$ iff it is created by $m_1$ or $m_2$.

However, according to \cite{taentzer2014fundamental}, model modifications can be in conflict in two cases: (i) insert-delete conflict and (ii) delete-delete conflict. Taentzer et al. state that only (i), where one modification creates an edge connected to a node deleted by the other modification, is an actual conflict, which has to be resolved to create a correct merge result. In this case, the merge result may deviate from the default case. Such conflicts will be reported by $\mathop{mcheck}(\gspan{M_c}{K_i}{M_i}, \gspan{M_c}{K_j}{M_j})$ in the form $(e, v)$, where $e$ is an edge created by one of the modifications and $v$ is a node deleted by the other modification.

For a correct model versioning $\Delta^{M_{\{1, \dots, n\}}}$, we say that two sequences of model modifications $M_c \Rightarrow^* M_i$ and $M_c \Rightarrow^* M_j$ are in conflict iff their corresponding maximally preserving model modifications \gspan{M_c}{K_{c, i}}{M_i} and \gspan{M_c}{K_{c, j}}{M_j} are in conflict. In this case, we also say that $M_i$ and $M_j$ are in conflict for the common predecessor $M_c$.

Insert-delete conflicts can be resolved by equipping the $\mathop{merge}$ operation with a manual or automatic strategy for conflict resolution. We consider such a strategy valid if it decides for each conflict whether to either revert the edge creation or the node deletion and always produces a proper merged graph. The approach in \cite{taentzer2014fundamental} effectively proposes an automatic strategy that favors insertion over deletion in order to preserve as many model elements as possible. Therefore, it reverts any deletions of nodes that would lead to insert-delete conflicts.

In contrast, a strategy for conflict resolution may favor deletion over insertion by reverting any creations of edges that would lead to insert-delete conflicts. Specifically, for model modifications $m_1 = \gspan{M_c}{K_i}{M_i}$ and $m_2 = \gspan{M_c}{K_j}{M_j}$, the model modification $m_{min} = \mathop{merge^{min}}(m_1, m_2)$, with $\mathop{merge^{min}}$ a merge operation equipped with this strategy, only creates an edge created by $m_1$ or $m_2$ if neither its source nor target is deleted by the other modification.

If all well-formedness conditions are specified by simple violation patterns, $m_{min}$ also yields a model where all well-formedness violations are also present in the merge result for any other conflict resolution strategy:

\begin{theorem} \label{the:merge_min}
For two model modifications $m_1 = \gspan{M_c}{K_i}{M_i}$ and $m_2 = \gspan{M_c}{K_j}{M_j}$ and a well-formedness constraint $\phi$ with violation pattern $Q$, it holds that
$$
\mathop{pcheck}(\mathop{merge_G^{min}}(m_1, m_2), \phi) = \bigcap_{\mathit{str} \in S} \mathop{pcheck}(\mathop{merge_G^{str}}(m_1, m_2), \phi)
,$$
with $S$ the set of all valid conflict resolution strategies.
\begin{proof} (Sketch)
Follows directly from the fact that $\mathop{merge}_G^{min}(m_1, m_2)$ is the smallest common subgraph of all graphs produced by the operation $\mathop{merge}$ for any valid conflict resolution strategy. \qed
\end{proof}
\end{theorem}

If there are no conflicts in the merged model operations, the $\mathop{merge}$ operation produces the same result regardless of the chosen strategy for conflict resolution.

For a correct model versioning, two model versions $M_i$ and $M_j$, and the set of versions $P = \mathop{pre}(i) \cap \mathop{pre}(j)$, we define the function

\begin{equation*}
\mathop{pre^{C}}(i, j) = 
\begin{cases}
	\emptyset & M_i \in \mathop{pre}(j) \vee M_j \in \mathop{pre}(i) \\
	\{M_c \in P\,|\,\forall M_x \in P: M_c \notin \mathop{pre}(x)\} & \textrm{otherwise}
\end{cases},
\end{equation*}

which returns the set of latest common predecessors of $M_i$ and $M_j$. Note that our definition of $\mathit{pre}^C$ corresponds to the definition of a best common ancestor in conventional version control systems such as Git \cite{git}, which is used to compute the base for three-way merges in these systems.

\begin{figure}
\centering
\includegraphics[width=\textwidth]{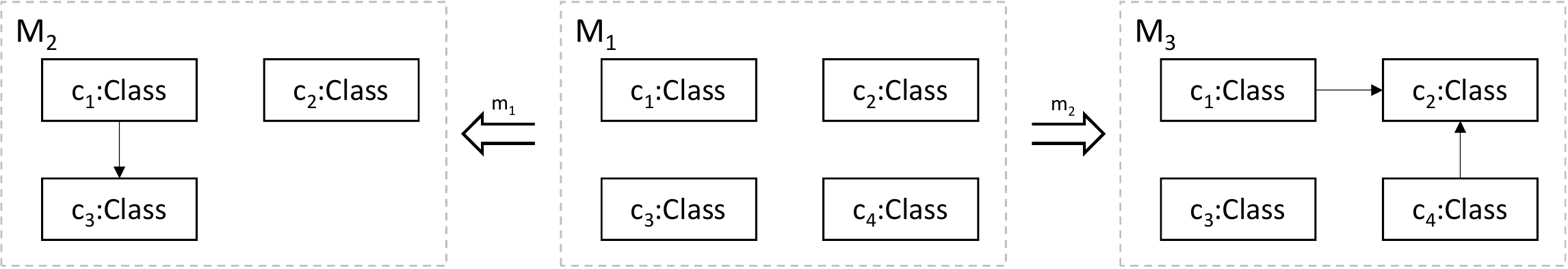}
\caption{Example model versioning} \label{fig:example_versioning}
\end{figure}

Figure \ref{fig:example_versioning} shows an exemplary model versioning based on the graph $M_1$ from Figure \ref{fig:example_graphs}. The initial graph $M_\alpha = M_1$ contains four classes. The modification $m_1$ (not to be confused with a morphism) to $M_2$ creates a superclass edge from $c_1$ to $c_3$ and deletes the node $c_4$. The modification $m_2$ to graph $M_3$ creates superclass edges from $c_1$ to $c_2$ and from $c_4$ to $c_2$. There is an insert-delete conflict between the two modifications, since the modification to $M_2$ deletes a node that is needed as the source of an edge created by the modification to $M_3$. Furthermore, the result of the merge of the two modifications would violate the well-formedness constraint with the violation pattern $Q$ from Figure \ref{fig:example_graphs}, since without additional modifications, the node $c_1$ would have two outgoing superclass edges.

\section{Multi-Version Models as Typed Graphs} \label{sec:multi-version_models}

%

A correct model versioning $\Delta^{M_{\{1, \dots, n\}}}$ with model versions $M_{\{1, \dots, n\}}$ conforming to a type graph $TM$ can be represented by a multi-version model in the form of a single graph that is typed over an adapted type graph.

The adapted type graph $TM_{mv}$ contains a node for each node and edge in $TM$. It also contains edges connecting each node in $TM_{mv}$ that represents an edge in $TM$ to the nodes representing the edge's source and target in $TM$. This yields a bijective function $\mathit{corr}_{mv}: V^{TM} \cup E^{TM} \rightarrow V^{TM_{mv}}$, which maps elements from $TM$ to the corresponding node in $TM_{mv}$, and two bijective functions $corr^s_{mv}, corr^t_{mv}: E^{TM} \rightarrow E^{TM_{mv}}$ mapping edges from $TM$ to the edges in $TM_{mv}$ encoding the source and target relation in $TM$. In addition, $TM_{mv}$ contains a node $\mathit{version}$, an edge $suc$ with source and target $\mathit{version}$, and two edges $cv_{v}$ and $dv_{v}$ from each other node $v \in V^{TM_{mv}}$ to the $version$ node.

A multi-version model $\mathit{MVM}$ for $\Delta^{M_{\{1, \dots, n\}}}$ is then constructed by an operation $\mathop{comb}$ as follows: 
A subgraph $P_{mv}^M$ encodes structural information about all model versions and is constructed by translating $P^M = \bigcup_{M_i \in M_{\{1, \dots, n\}}} M_i$ to conform to $TM_{mv}$ using an operation $\mathop{trans_{mv}}$. Since source and target functions are invariant in a correct model versioning, $P^M$ is well-defined.

For each $v \in v^{P^M}$, $\mathop{trans_{mv}}$ creates a node of type $\mathit{corr}_{mv}(v)$ in $V^{P_{mv}^M}$. For each $e \in E^{P^M}$, a node of type $\mathit{corr}_{mv}(e)$ is created. This yields a bijection $\mathit{origin}: P_{mv}^M \rightarrow P^M$ mapping translated elements to their original representation.

In addition, for each edge $e \in E^{P^M}$, an edge of type $corr^s_{mv}(e)$ with source $\mathit{origin}^{-1}(e)$ and target $\mathit{origin}^{-1}(s^{P^M}(e))$ and an edge of type $corr^t_{mv}(e)$ with source $\mathit{origin}^{-1}(e)$ and target $\mathit{origin}^{-1}(t^{P^M}(e))$ are created in $E^{P^M_{mv}}$. Since edge sources and targets are invariant, the corresponding node $v_e = \mathit{origin}^{-1}(e)$ in the end has exactly one edge of type $corr^s_{mv}(e)$ and one of type $corr^t_{mv}(e)$. We thus have two functions $s_{mv} : \mathit{origin}^{-1}(E^{P^M}) \rightarrow E^{P^M_{mv}}$ respectively $t_{mv} : \mathit{origin}^{-1}(E^{P^M}) \rightarrow E^{P^M_{mv}}$ encoding these mappings.

Another, distinct subgraph $P_{mv}^V$ contains versioning information and is constructed as follows: For each $M_i \in M_{\{1, \dots, n\}}$, $P_{mv}^V$ contains a corresponding node of type $version$. For each $\gspan{M_i}{K}{M_j} \in \Delta^{M_{\{1, \dots, n\}}}$, $P_{mv}^V$ contains an edge of type $suc$ from the node representing $M_i$ to the node representing $M_j$.

For each modification \gspan{M_i}{K}{M_j}, a $cv$-edge with the node corresponding to $M_j$ as its target is added to all nodes corresponding to elements created by the modification. A $dv$-edge with the node corresponding to $M_j$ as its target is added to all nodes corresponding to elements deleted by the modification. Additionally, a $cv$ edge with the node corresponding to the initial version $M_\alpha$ as its target is added to all nodes corresponding to elements in $M_\alpha$.

Since attributes can be encoded by dedicated nodes and assignment edges \cite{Heckel+2002}, the construction can be performed analogously for attributed graphs.

For $v \in P_{mv}^M$ and $M_i \in M_{\{1, \dots, n\}}$, we say that $v$ is \emph{mv-present} in $M_i$, iff for a node $m_{cv}$ connected to $v$ via a $cv$ edge, there exists a path from $m_{cv}$ to the node representing $M_i$ via $suc$ edges that does not go through a node connected to $v$ via a $dv$ edge. We denote the set of versions where $v$ is mv-present by $p(v)$.

A model version $M_i$ can then be derived from $\mathit{MVM}$ via an operation $\mathop{proj}$ as follows: 
Collect all nodes $V_p = \{v_p \in V^{P_{mv}^M} | M_i \in p(v_p) \}$, that is, all nodes that are mv-present in $M_i$, and translate the induced subgraph into the single-version model $M_i$ with $V^{M_i} = \{\mathit{origin}(v_v) | v_v \in V^{MVM} \wedge \mathit{corr}_{mv}^{-1}(type^{MVM}(v_v)) \in V^{TM}\}$, $E^{M_i} = \{\mathit{origin}(v_e) | v_e \in V^{MVM} \wedge \mathit{corr}_{mv}^{-1}(type^{MVM}(v_e)) \in E^{TM}\}$,  $s^{M_i} = \mathit{origin} \circ t^{MVM} \circ s_{mv} \circ \mathit{origin}^{-1}$, and $t^{M_i} = \mathit{origin} \circ t^{MVM} \circ t_{mv} \circ \mathit{origin}^{-1}$.

\subsubsection{Correctness}

\begin{theorem} \label{the:proj}
For a correct model versioning $\Delta^{M_{\{1, \dots, n\}}}$ holds concerning $\mathop{comb}$ and $\mathop{proj}$:
$$
  \forall i \in \{1, \dots, n\}:
    M_i  = \mathop{proj}(\mathop{comb}(\Delta^{M_{\{1, \dots, n\}}}),i)
.$$
\begin{proof} (Sketch)
Any element in a version $M_i$ has a corresponding node $v$ in $\mathop{comb}(\Delta^{M_{\{1, \dots, n\}}})$. By construction, $v$ is connected to a node corresponding to some version $M_j$ via a $cv$ edge, for which there exists a path of $\mathit{suc}$ edges to the node corresponding to $M_i$. That path does not go through a node connected to $v$ by a $dv$ edge. $v$ is thus mv-present in $M_i$ and hence contained in the projection.

Inclusion of elements in the opposite direction can be shown analogously. Because edge sources and targets are invariant over all graphs, the edges in $\mathop{comb}(M_1, \dots, M_n)$ correctly encode the source and target functions by construction. Thus, $\forall i \in \{1, \dots, n\}: M_i  = \mathop{proj}(\mathop{comb}(M_1, \dots, M_n),i)$. \qed
\end{proof}
\end{theorem}

A maximally preserving model modification \gspan{M_i}{K}{M_j} with $M_i, M_j \in M_{\{1, \dots, n\}}$ (and thus any model modification in $\Delta^{M_{\{1, \dots, n\}}}$) can be derived from $\mathit{MVM}$ via $\mathop{proj^\Delta}$ as follows: $M_i$ and $M_j$ can be derived via the operation $\mathop{proj}$. $K$ is then the graph containing all elements from $M_i \cap M_j$, with $s^K$ and $t^K$ uniquely defined by the corresponding functions from $M_i$ and $M_j$ and partial identities as morphisms into $M_i$ and $M_j$.

\begin{theorem}
For a correct model versioning $\Delta^M_{\{1, \dots, n\}}$, concerning $\mathop{comb}$ and $\mathop{proj^\Delta}$, it holds that:
$$
  \forall M_i, M_j \in M_{\{1, \dots, n\}}:
    m_{i, j} = \mathop{proj^\Delta}(\mathop{comb}(\Delta^{M_{\{1, \dots, n\}}}), i, j)
,$$
with $m_{i, j}$ the maximally preserving model modification from $M_i$ to $M_j$.

\begin{proof}
Follows trivially from Theorem \ref{the:proj} and the definition of the maximally preserving model modification \gspan{M_i}{K_{i,j}}{M_j}. \qed
\end{proof}
\end{theorem}

\begin{figure}
\centering
\includegraphics[width=\textwidth]{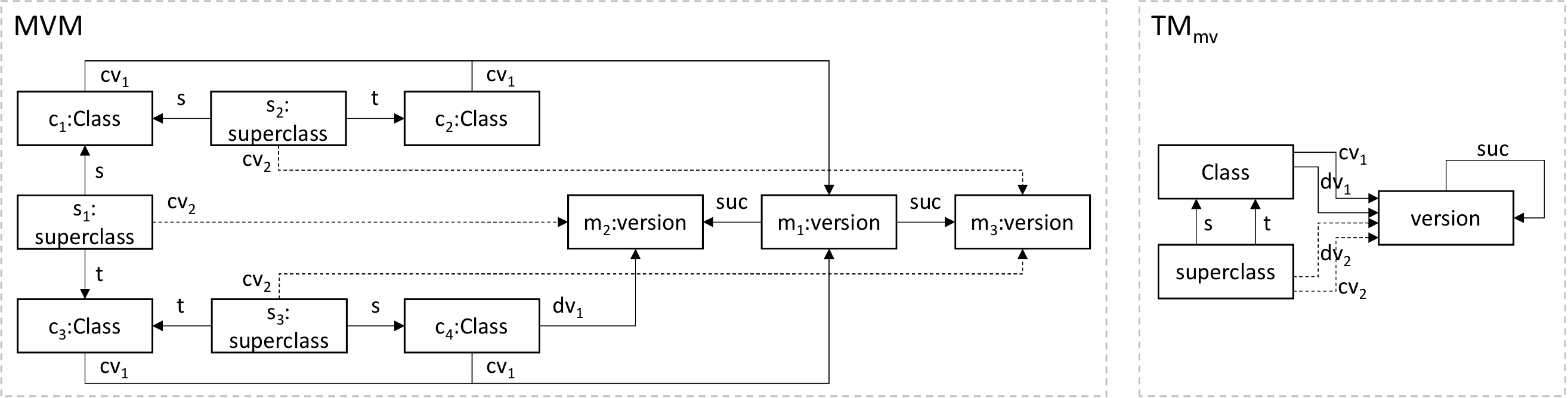}
\caption{Multi-version model and adapted type graph for the versioning in Figure \ref{fig:example_versioning}} \label{fig:example_mvm}
\end{figure}

Figure \ref{fig:example_mvm} visualizes the multi-version model $MVM$ constructed for the example versioning in Figure \ref{fig:example_versioning} and the associated adapted type graph $TM_{mv}$. $MVM$ contains a node for each node and edge in the models of the example versioning, one node of type $version$ for each of the graphs $M_1$, $M_2$, and $M_3$, and appropriate edges as created by $\mathop{comb}$.

%
\subsection{Directly Checking Well-Formedness for Multi-Version Models}

We can use a multi-version model to directly find all well-formedness violations in all individual versions via an operation $\mathop{pcheck_{mv}}$. For a multi-version model $\mathit{MVM}$ with a bijective mapping into a union of original model versions $\mathit{origin}_M$ and a well-formedness constraint $\phi$ with associated violation pattern $Q$, $\mathop{pcheck_{mv}}(\mathit{MVM}, \phi)$ works as follows:

First, the graph $Q$ typed over the original type graph is translated into a corresponding graph $Q_{mv}$ typed over the adapted type graph using $\mathop{trans_{mv}}$. This yields a bijective mapping $\mathit{origin_Q} : Q_{mv} \rightarrow Q$.

Then, all matches for $Q_{mv}$ in $\mathit{MVM}$ are found. For each such match $m_{mv}$, $\mathop{pcheck_{mv}}$ computes all versions for which all vertices in the image of the match are mv-present by $P = \bigcap_{v \in V^{Q_{mv}}} p(m_{mv}(v))$. If $P \neq \emptyset$, the match into the original model versions $m = \mathit{origin_{M}} \circ m_{mv} \circ \mathit{origin_{Q}^{-1}}$ is constructed and reported as a violation in all versions in $P$.

\subsubsection{Correctness}

\begin{theorem} \label{the:pcheck}
For a well-formedness constraint $\phi$ with violation pattern $Q$, a correct model versioning $\Delta^{M_{\{1, \dots, n\}}}$, and $\mathit{MVM} = \mathop{comb}(\Delta^M_{\{1, \dots, n\}})$ holds:
$$
    \mathop{pcheck_{mv}}(\mathit{MVM},\phi) 
  = 
    \biguplus_{i \in \{ 1, \dots, n\}}
     \{(i, m)| m\in \mathop{pcheck}(\mathop{proj}(\mathit{MVM},i),\phi)\}
  .
$$

\begin{proof} (Sketch)
A match $m: Q \rightarrow M_i$ for any version $M_i$ has one corresponding match $m_{mv}$ with $m = \mathit{origin_{M}} \circ m_{mv} \circ \mathit{origin_{Q}^{-1}}$, where edges created by $\mathop{trans}_{mv}$ ensure correct connectivity. $P = \bigcap_{v \in V^{Q_{mv}}} p(m_{mv}(v))$ contains exactly the versions containing all elements in $m(Q)$. This yields the stated equality. \qed
\end{proof}
\end{theorem}

\subsubsection{Complexity} \label{sec:pcheck_mv_complexity}

The effort for searching all versions $M_{\{1, \dots, n\}}$ of some model versioning $\Delta^{M_{\{1, \dots, n\}}}$ for a pattern $Q$ using $\mathop{pcheck}$ is in $O(\sum_{M_i \in M_{\{1, \dots, n\}}} C(M_i, Q))$, with $C(M_i, Q)$ the effort for finding all matches of $Q$ into $M_i$.

$P_{mv}^M = \mathop{trans}_{mv}(P^M)$ and $Q_{mv} = \mathop{trans}_{mv}(Q)$ are only different encodings of $P^M = \bigcup_{M_i \in M_{\{1, \dots, n\}}} M_i$ and $Q$. Considering computation of the mv-present predicate, the effort for $\mathop{pcheck_{mv}}$ is hence in $O(C(\bigcup_{M_i \in M_{\{1, \dots, n\}}} M_i, Q) + X \cdot |V^{Q_{mv}}| \cdot |\Delta^{M_{1, \dots, n}}|)$, with $X$ the number of matches for  $Q_{mv}$ into $P_{mv}^M$.

\subsubsection{Discussion}

If many elements are shared between individual versions and modifications only perform few changes, the size of the union of all model versions will be small compared to the sum of the sizes of all individual versions. If pattern matching is efficient with respect to the size of the considered model, pattern matching over the union of all model versions will then likely require less effort than matching over each individual version. If the number of matches for violation patterns is low, the associated checks performed by $\mathop{pcheck_{mv}}$ will likely be more efficient than the pattern matching over the individual versions.

Overall, $\mathop{pcheck_{mv}}$ will thus likely be more efficient than using $\mathop{pcheck}$ in scenarios where pattern matching is efficient, the number of changes between versions is low, and the number of violations in the union of versions is low.


\section{Directly Checking Merge Results for Multi-Version Models} \label{sec:checking_merge_results}

We can consider multi-version models to directly detect whether 
(a) merge conflicts exist for any valid pair of encoded model modifications via an operation $\mathop{mcheck_{mv}}$ and
(b) any resulting merged model is ill-formed via an operation $\mathop{pcheck^m_{mv}}$, 
where a pair of model modifications \gspan{M_c}{K_i}{M_i} and \gspan{M_c}{K_j}{M_j} is valid iff $M_c \in \mathit{pre}^C(M_i, M_j)$.

%
\subsection{Directly Checking for Merge Conflicts}

$\mathop{mcheck_{mv}}$ can be realized for a multi-version model $\mathit{MVM} = \mathop{comb}(\Delta^{M_{\{1, \dots, n\}}})$ as follows: First, the operation collects all nodes in $\mathit{MVM}$ representing edges that are created by some model modification. This means all nodes $v_e \in V^{MVM}$ where $\mathit{corr}_{mv}^{-1}(type^{MVM}(v)) \in E^{TM}$ connected to a node $m_x$ via a $cv$ edge, where $m_x$ does not correspond to $M_\alpha$ and with $TM$ the original type graph. Then, for each node $v_e$, we compute the set of versions $P = p(v_e)$ where it is mv-present. If $P \neq p(v_s)$, where $v_s = s^{\mathit{MVM}}(s_{mv}(v_e))$, we then compute a set of versions $D$ that correspond to nodes reachable via $\mathit{suc}$ edges from a node connected to $v_s$ via a $dv$ edge without going through nodes connected to $v_s$ via a $cv$ edge.

Afterwards, for each pair of versions $M_i \in P$ and $M_j \in D$, we check for each latest common predecessor $M_c \in \mathit{pre}^C(i, j)$ whether $M_c \in p(v_s) \wedge M_c \notin P$. For any triplet of versions $(i, j, c)$ where this is the case, the edge $\mathit{origin}(v_e)$ is then in an insert-delete conflict with its source. To facilitate formalization, this conflict is reported in the normalized form $(\mathop{min}(i, j), \mathop{max}(i, j), c, (\mathit{origin}(v_e), \mathit{origin}(v_s)))$. Insert-delete conflicts with the edge's target are computed analogously.

\subsubsection{Correctness}

\begin{theorem}

For a model versioning $\Delta^{M_{\{1, \dots, n\}}}$ and the associated multi-version model $\mathit{MVM} = \mathop{comb}(\Delta^M_{\{1, \dots, n\}})$ holds:

\begin{equation*}
\mathop{mcheck_{mv}}(\mathit{MVM}) =\biguplus_{(i,j,c) \in Y} \{(i, j, c, m) | m \in \mathop{mcheck}(m_{c, i}, m_{c, j})\},
\end{equation*}
where $Y = \{(i, j, c)\,|\,i, j \in \{ 1, \dots, n\}:  i < j, c \in \{c | M_c \in \mathop{pre^C}(i, j)\}\}$ and with $m_{c, i} = \mathop{proj^\Delta}(\mathit{MVM},c, i)$ and $m_{c, j} = \mathop{proj^\Delta}(\mathit{MVM},c, j)$.

\begin{proof} (Sketch)
The collected nodes representing edges correspond to a superset of edges that may be involved in a conflict. The construction of the sets $P$ and $D$ for a collected node $v_e$ ensures that any pair of versions where one may create $e = \mathit{origin}(v_e)$ and the other may delete the source (or target) of $e$ is considered. The condition checked for each common predecessor of a version pair then yields exactly the triplets of versions where $e$ is part of an insert-delete conflict. Because of the normalization of the results of $\mathop{mcheck_{mv}}$, we have the stated equality. \qed
\end{proof}
\end{theorem}

\subsubsection{Complexity}

The function $\mathit{pre}^C_{mv}$ can be precomputed in $O(|M_{\{1,\dots,n\}}|^4)$.

Since information about creation and deletion of elements is not explicitly available in a na\"ive representation, finding all insert-delete conflicts between two model modifications via $\mathop{mcheck}$ has to be done by checking for each edge in either modification's resulting model whether it is created by that modification and its source or target is deleted by the other modification. Since there may exist up to $O(|M_{\{1,\dots,n\}}|^3)$ possible merges in a model versioning, in the worst case, this implies effort in $O(|M_{\{1,\dots,n\}}|^4 + |E^{M_{max}}| \cdot |M_{\{1,\dots,n\}}|^3)$, where $|E^{M_{max}}|$ is the maximum number of edges present in a single model version.

Created edges can be retrieved efficiently from a multi-version model given appropriate data structures. Computing and checking the required version sets takes $O(|M_{\{1,\dots,n\}}|^3)$ steps per edge. Therefore, the overall computational complexity of $\mathop{mcheck}_{mv}$ is in $O(|M_{\{1,\dots,n\}}|^4 + \Delta_+ \cdot |M_{\{1,\dots,n\}}|^3)$, where $\Delta_+$ is the overall number of elements created in the model versioning.

\subsubsection{Discussion}

The efficiency of $\mathop{mcheck}_{mv}$ compared to using $\mathop{mcheck}$ mostly depends on the number of edges created by some model modification compared to the number of edges in the individual versions. If most edges are present in the original model version and are shared between many model versions, $\mathop{mcheck}_{mv}$ will be more efficient. Otherwise, $\mathop{mcheck}_{mv}$ will not achieve a significant improvement and might even perform worse than the operation based on $\mathop{mcheck}$.

Version control systems such as Git typically select a single latest common predecessor as the base for a three way merge\cite{git}. Using a corresponding partial function $\mathit{pre}^C_1 : \mathbb{N} \times \mathbb{N} \rightarrow M_{\{1,\dots,n\}}$ with $\mathit{pre}^C_1(i, j) \in \mathit{pre}^C(i, j)$ if $\mathit{pre}^C(i, j) \neq \emptyset$ and $\mathit{pre}^C_1(i, j) = \perp$ to select a single latest common predecessor of two versions $i$ and $j$ rather than $\mathit{pre}^C$ in $\mathop{mcheck_{mv}}$, by the same logic as used in the proof of correctness, we instead have an analogous equality for $\mathit{pre}^C_1$. Disregarding the computational effort for precomputing $\mathit{pre}^C_1$, replacing $\mathit{pre}^C$ by $\mathit{pre}^C_1$ reduces the remaining computational complexity of $\mathop{mcheck_{mv}}$ to $O(\Delta_+ \cdot |M_{\{1,\dots,n\}}|^2)$.

%
\subsection{Directly Checking Well-Formedness for Merge Results}

To find all violations of a well-formedness constraint $\phi$ characterized by a pattern $Q$ via $\mathop{pcheck^m_{mv}}$ in merge results of a multi-version model $MVM$, we first translate $Q$ into $Q_{mv} = \mathop{trans}_{mv}$. We then find all matches for $Q_{mv}$ in $MVM$.

For a match $m_{mv}$ for $Q_{mv}$, we determine the set of versions $P_{v} = p(v)$ for each $v \in m_{mv}(V^{Q_{mv}})$. For each pair of versions $M_i \in \argmin_{P \in \{ p(v) | v \in m_{mv}(V^{Q_{mv}}\}} |P|$ and $M_j \in \bigcup_{v \in V^{Q_{mv}}} p(v)$, we check whether $\forall v \in m_{mv}(V^{Q_{mv}}): M_i \in p(v) \vee M_j \in p(v)$. We then check for each latest common predecessor $M_c \in \mathit{pre}^C(i, j)$ if for all $v \in V^{Q_{mv}}$, it holds that $v \in V^{M_c} \rightarrow (v \in V^{M_i} \wedge v \in V^{M_j})$, that is, $v$ is not deleted in $M_i$ or $M_j$. If this is the case, the match $m$ into $\bigcup_{M_x \in M_{\{1,\dots,n\}}} M_x$ corresponding to $m_{mv}$ represents a violation in $\mathop{merge}^{min}(\gspan{M_c}{K_i}{M_i}, \gspan{M_c}{K_j}{M_j})$. We report results in the normalized form $(\mathop{min}(i, j), \mathop{max}(i, j), c, m)$.

\subsubsection{Correctness}

\begin{theorem} \label{the:pcheck_merge}
For a well-formedness constraint $\phi$, a correct model versioning $\Delta^{M_{\{1, \dots, n\}}}$, and the multi-version model $\mathit{MVM} = \mathop{comb}(\Delta^M_{\{1, \dots, n\}})$ holds:
\begin{equation*}
\mathop{pcheck^m_{mv}}(\mathit{MVM},\phi) = \biguplus_{(i, j, c) \in Y} \{(i, j, c, m) | m \in \mathop{pcheck}(M^{min}_{i, j, c}, \phi)\},
\end{equation*}
where $Y = \{(i, j, c)\,|\,i, j \in \{ 1, \dots, n\}:  i < j, c \in \{c | M_c \in \mathop{pre^C}(i, j)\}\}$ and $M^{min}_{i, j, c} = \mathop{merge_G^{min}}(\mathop{proj^\Delta}(\mathit{MVM},c, i),\mathop{proj^\Delta}(\mathit{MVM},c, j))$.

\begin{proof} (Sketch)
For two versions $M_i, M_j$ with latest common predecessor $M_c$, a match $m: Q \rightarrow \mathop{merge_G^{min}}(\mathop{proj^\Delta}(\mathit{MVM},c, i),\mathop{proj^\Delta}(\mathit{MVM},c, j))$  has one corresponding match $m_{mv}: \mathop{trans}_{mv}(Q) \rightarrow \mathit{MVM}$ by construction, where the edges created by $\mathop{trans}_{mv}$ ensure the correct connectivity. The set of version pairs considered by $\mathop{pcheck^m_{mv}}$ contains all version pairs such that each matched element is contained in at least one of the versions. The condition checked for every latest common predecessor ensures that only version triplets are reported where the merge result also contains all matched elements if there are no merge conflicts. Since $\mathop{merge^{min}}$ resolves conflicts by prioritizing deletion and, as ensured by the check, no matched node is deleted by the merge, conflict resolution cannot invalidate the match or create new matches. We thus have the stated equality. \qed
\end{proof}
\end{theorem}

By Theorem \ref{the:merge_min} and Theorem \ref{the:pcheck_merge}, we also have that $\mathop{pcheck^m_{mv}}$ yields the set of violations that cannot be avoided by any conflict resolution strategy:

\begin{corollary}
For a well-formedness constraint $\phi$, a correct model versioning $\Delta^{M_{\{1, \dots, n\}}}$, and the multi-version model $\mathit{MVM} = \mathop{comb}(\Delta^{M_{\{1, \dots, n\}}})$ holds:
\begin{equation*}
\mathop{pcheck^m_{mv}}(\mathit{MVM},\phi) = \biguplus_{(i, j, c) \in Y} \bigcap_{\mathit{str} \in S} \{(i, j, c, m) | m \in \mathop{mcheck}(M^{str}_{i, j, c}, \phi)\},
\end{equation*}
where $Y = \{(i, j, c)\,|\,i, j \in \{ 1, \dots, n\}:  i < j, c \in \{c | M_c \in \mathop{pre^C}(i, j)\}\}$ and $M^{str}_{i, j, c} = \mathop{merge_G^{str}}(\mathop{proj^\Delta}(\mathit{MVM},c, i),\mathop{proj^\Delta}(\mathit{MVM},c, j))$, and with $S$ the set of all valid conflict resolution strategies.
\end{corollary}

\subsubsection{Complexity}

The function $\mathit{pre}^C_{mv}$ can be precomputed in $O(|M_{\{1,\dots,n\}}|^4)$.

With $C(M_i, Q)$ the effort for finding all matches of $Q$ into $M_i$, finding violations characterized by a pattern $Q$ in all results of a set of possible merges $Y$ using $\mathop{pcheck}$ takes effort in $O(O(|M_{\{1,\dots,n\}}|^4 +\sum_{(m_1, m_2) \in Y} C(\mathop{merge^{min}_G}(m_1, m_2), Q))$.

The computation and checking of version triplets for a match in $\mathop{pcheck^m_{mv}}$ takes effort in $O(|M_{\{1, \dots, n\}}|^3)$. For $X$ matches for $Q_{mv}$, the effort for $\mathop{pcheck^m_{mv}}$ is thus in $O(|M_{\{1,\dots,n\}}|^4 + C(\bigcup_{M_i \in M_{\{1, \dots, n\}}} M_i, Q) + X \cdot |V^{Q_{mv}}| \cdot |M_{\{1, \dots, n\}}|^3)$.

\subsubsection{Discussion}

By the same argumentation as for $\mathop{pcheck_{mv}}$, $\mathop{pcheck^m_{mv}}$ will likely be more efficient than the corresponding operation using $\mathop{pcheck}$ in scenarios where pattern matching is efficient, the number of changes between versions is low, and the number of violations in the union of model versions is low.

Using some partial function $\mathit{pre}^C_1 : \mathbb{N} \times \mathbb{N} \rightarrow M_{\{1,\dots,n\}}$ to select a single latest common predecessor rather than $\mathit{pre}^C$ in $\mathop{pcheck^m_{mv}}$, by the same logic as in the proof of correctness, we have an analogous equality for $\mathit{pre}^C_1$. Disregarding the effort for precomputing $\mathit{pre}^C_1$, replacing $\mathit{pre}^C$ by $\mathit{pre}^C_1$ reduces the remaining complexity of $\mathop{pcheck^m_{mv}}$ to $O(C(\bigcup_{M_i \in M_{\{1, \dots, n\}}} M_i, Q) + X \cdot |V^{Q_{mv}}| \cdot |M_{\{1, \dots, n\}}|^2)$.


\section{Evaluation}

For an initial empirical evaluation of the performance and scalability of the presented operations, we experiment with an application scenario from the software development domain. Therefore, we extract abstract syntax graphs from a small previous research project (\textbf{rete}) and a larger open source project (\textbf{henshin} \cite{arendt2010henshin}) written in Java using the EMF-based \cite{emf} MoDisco tool \cite{bruneliere2010modisco}. We store the extracted models in a graph format and fold each of the projects into a multi-version model, using a mapping strategy based on hierarchy and element names.

We then run implementations of the presented operations for conflict detection and well-formedness checking based on multi-version models (\textbf{MVM}) and baseline implementations using corresponding single-version models (\textbf{SVM}).\footnote{All experiments were executed on a Linux SMP Debian 4.19.67-2 machine with Intel Xeon E5-2630 CPU (2.3\,GHz clock rate) and 386\,GB system memory running OpenJDK version 1.8.0\_242. Reported execution times correspond to the arithmetic mean of at least five runs of the respective experiment. Our implementation and datasets are available under \url{https://github.com/hpi-sam/multi-version-models}} We consider three well-formedness constraints: uniqueness of a class's superclass, uniqueness of a method's return type, and consistency of an overriden method's return type. We employ our own EMF-based tool \cite{giese2009improved} for pattern matching.

\begin{figure}
\centering
\includegraphics[width=\textwidth]{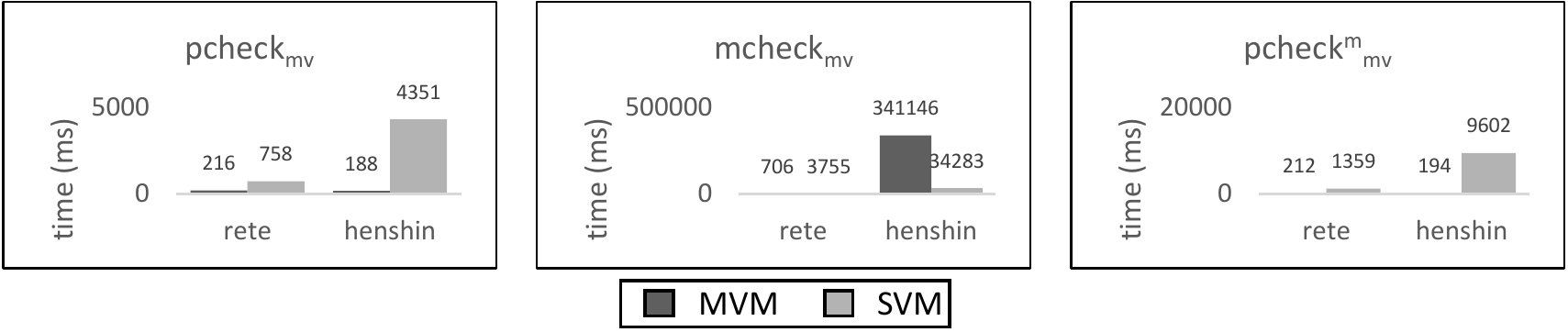}
\caption{Measurement results for $\mathop{pcheck_{mv}}$, $\mathop{mcheck_{mv}}$, and $\mathop{pcheck^m_{mv}}$} \label{fig:measurements}
\end{figure}

Figure \ref{fig:measurements} shows the measured execution times for the operations $\mathop{pcheck_{mv}}$, $\mathop{mcheck_{mv}}$, and $\mathop{pcheck^m_{mv}}$ and related single-version-model-based operations over the example models. The execution times for $\mathop{pcheck_{mv}}$ and $\mathop{pcheck^m_{mv}}$ correspond to the combined pattern matching time for all considered well-formedness constraints. All reported times exclude the time for computing any merge results required by SVM and the time required to precompute the $\mathit{pre}^C$ function, since it is required by both the MVM and the SVM implementation. Precomputing $\mathit{pre}^C$ took about 5\,ms for the smaller project and about 3.5\,s for the larger project.

For the tasks related to well-formedness checking, the MVM variant performs better (up to factor 50) than SVM. Since there are only few to no matches for the violation patterns of the considered constraints, the MVM implementation only performs few of the potentially expensive checks over the version graph, while avoiding most of the redundancy in the pattern matching of SVM.

For conflict detection, MVM performs better than SVM for the smaller project (factor 5), but has a substantially higher execution time for the larger project (factor 10). The reason for the bad performance is that most edges are not present in the initial model version. In fact, the number of edges created throughout the model versioning is much higher than the number of edges in any individual version. Furthermore, in contrast to the solution using $\mathop{mcheck}$, the operation $\mathop{mcheck}_{mv}$ considers versions where the source or target of an edge is \emph{not} present. Due to the high number of versions in the project and because many elements are only present in few versions, this leads to the processing of large version sets, which deteriorates the performance of MVM in this scenario.

\subsubsection{Threats to Validity}

Unexpected JVM behavior poses a threat to internal validity, which we tried to mitigate by performing multiple runs of each experiment and profiling time spent on garbage collection. To address threats to external validity, we used real-world data and well-formedness constraints in our experiments. While we used our own tool for pattern matching, said tool has already been used in our previous works and has shown adequate performance \cite{giese2009improved}.

However, the example constraints are not representative and the folding of individual model versions extracted from source code may yield a larger-than-necessary multi-version model. Our results are thus not necessarily generalizable, but instead constitute an early conceptual evaluation of the presented approach.


\section{Related Work}

While most practical version control systems operate on text documents \cite{Mens2005}, versioning and merging of models has also been subject to extensive research.

There already exist several formal and semi-formal approaches to model merging, which compute the result of a three-way-merge of model modifications \cite{westfechtel2010formal,taentzer2014fundamental}. Notably, the approach by Taentzer et al. \cite{taentzer2014fundamental} represents a formally defined solution that works on the level of graphs, which is why for our approach, we build on their notion of model merging. In their work, Taentzer et al. also consider checking of well-formedness constraints by constructing a tentative merge result over which the check is executed. While this allows their approach to handle arbitrary constraints rather than just simple graph patterns, the check has to be executed for each individual merge.

Some approaches consider detection of merge conflicts \cite{kuster2009dependent} or model inconsistencies \cite{blanc2009incremental} based on the analysis of sequences of primitive changes. However, these approaches do not consider the case of multiple versions and pairwise merges and naturally do not employ a graph-based definition of inconsistencies.

For the more general problem of model versioning, both formal solutions \cite{rutle2009category,diskin2009model} and tool implementations \cite{murta2008towards,koegel2010emfstore} have been introduced. Similar to our approach, some of these techniques are based on a joint representation of multiple model versions \cite{murta2008towards,rutle2009category}. However, to the best of our knowledge, joint conflict detection or well-formedness checking for all merges at once is not considered.

Westfechtel and Greiner \cite{westfechtel2020extending} present a solution to configuration management for product lines, where several configurations are represented in a unified feature model. Information about the presence of elements in certain configurations is encoded in logic formulas and propagated along model transformations. While this approach bears some similarity to the encoding of different versions and collective well-formedness checking in our solution, the technique in \cite{westfechtel2020extending} focuses on the definition of product lines and hence does not consider merging.

Folding and joint querying of the temporal evolution of graphs has been studied in previous work of our group \cite{giese2019metric,sakizloglou2021incremental}. However, the aim of these solutions is the development of a temporal logic for graphs. The presented concepts are hence specific to sequences of graph modifications and do not consider merging.


\section{Conclusion}

In this paper, we have presented an approach for encoding a model versioning as a single typed graph. Based on this representation, we have introduced operations for finding merge conflicts and violations of well-formedness conditions in the form of graph patterns in the entire versioning and related merge results. We have conducted an initial empirical evaluation, which demonstrates potential benefits of the approach, but also highlights shortcomings in unfavorable scenarios.

In future work, we plan to address these shortcomings by studying how to compress the version graph or restrict the set of considered versions to those most relevant to users. We want to investigate how to lift our notion of well-formedness constraints to more expressive formalisms such as nested graph conditions and develop an incremental version of the approach. Finally, we will extend our empirical evaluation to better characterize our technique's performance.

\renewcommand{\doi}[1]{\href{http://doi.org/#1}{\url{http://doi.org/#1}}}
\bibliographystyle{llncs_template/splncs04}
\bibliography{mvm_conflicts}

\newpage

\appendix


\section{Technical Details}

\begin{theorem}
For a correct model versioning $\Delta^{M_{\{1, \dots, n\}}}$ holds concerning $\mathop{comb}$ and $\mathop{proj}$:
$$
  \forall i \in \{1, \dots, n\}:
    M_i  = \mathop{proj}(\mathop{comb}(\Delta^{M_{\{1, \dots, n\}}}),i)
.$$
\begin{proof}
For any version $M_i$ and any element $x \in M_i$, there is a version $M_j$, such that $x \in M_j$ with $M_j = M_\alpha$ or $\exists \gspan{M_x}{K}{M_j} \in \Delta^{M_{\{1, \dots, n\}}}: x \notin M_x \wedge x \in M_j$, that is, \gspan{M_x}{K}{M_j} creates $x$, and there exists a (potentially empty) sequence of model modifications $M_j \Rightarrow^* M_i$ that does not delete $x$ again.

Thus, by the construction of $\mathop{comb}$, there exists an element $x_{mv} \in \mathit{MVM}$, where $\mathit{MVM} = \mathop{comb}(M_1, \dots, M_n)$, which is connected to the node $m_j \in MVM$ that corresponds to $M_j$ via a $cv$ edge.

Because there exists a sequence of model modifications $M_j \Rightarrow^* M_i$ where $x$ is not deleted, there exists a path of $suc$ edges from $m_j$ to the node representing $M_i$ that does not go through any node connected to $x_{mv}$ via a $dv$ edge. Therefore, $x_{mv}$ is mv-present in $M_i$ for $\mathit{MVM}$. Hence, $x \in \mathop{proj}(\mathop{comb}(M_1, \dots, M_n),i)$.

Since all implications used for deductions are equivalences, we have that $x \in \mathop{proj}(\mathop{comb}(M_1, \dots, M_n),i) \Leftrightarrow x \in M_i$ for any element $x$ and model version $M_i$ and therefore the equality of vertex and edge sets of $M_i$ and the projection.

Because edge sources and targets are invariant over multiple graphs, the edges in $\mathop{comb}(M_1, \dots, M_n)$ correctly encode the source and target functions by construction. Thus, $\forall i \in \{1, \dots, n\}: M_i  = \mathop{proj}(\mathop{comb}(M_1, \dots, M_n),i)$. \qed
\end{proof}
\end{theorem}

\begin{theorem}
For a well-formedness constraint $\phi$, a correct model versioning $\Delta^{M_{\{1, \dots, n\}}}$, and associated multi-version model $\mathit{MVM} = \mathop{comb}(\Delta^M_{\{1, \dots, n\}})$ holds:
$$
    \mathop{pcheck_{mv}}(\mathit{MVM},\phi) 
  = 
    \biguplus_{i \in \{ 1, \dots, n\}}
     \{(i, m)| m\in \mathop{pcheck}(\mathop{proj}(\mathit{MVM},i),\phi)\}
  .
$$

\begin{proof}
For a tuple $(i, m)$ with match $m: Q \rightarrow \bigcup_{M_i \in M_{\{1, \dots, n\}}} M_i$ to be reported in the result of $\mathop{pcheck_{mv}}(M_{\{1,\dots,n\}},\phi)$, there has to exist a match $m_{mv}: Q_{mv} \rightarrow \mathit{MVM}$, where $Q_{mv} = \mathop{trans}_{mv}(Q)$, such that $m = \mathop{trans_{mv}}^m(m_{mv})$ and $\bigcap_{v \in V^{Q_{mv}}}p(m_{mv}(v)) \neq \emptyset$.

There hence exists some version $M_i \in M_{\{1, \dots, n\}}$ such that $\forall v \in V^{Q_{mv}}: M_i \in p(m_{mv}(v))$. Therefore, $\forall v \in V^{Q_{mv}}: \mathit{origin_M}(m_{mv}(v)) \in \mathop{proj}(M_{\{1,\dots,n\}},i)$.

Then, $m$ is a valid monomorphism into $M_i = \mathop{proj}(M_{\{1,\dots,n\}},i)$, since (i) all required elements are present and (ii) by construction of $\mathit{MVM}$, $Q_{mv}$, the edges in $Q_{mv}$ ensure that $\forall e \in E^Q: s^{M_i}(m(e)) = m(s^Q(e)) \wedge t^{M_i}(m(e)) = m(t^Q(e))$, and the $\mathit{origin}$ function is bijective. Therefore, it holds that $(i, m) \in \biguplus_{i \in \{ 1, \dots, n\}} \mathop{pcheck}(\mathop{proj}(M_{\{1,\dots,n\}},i),\phi)$.

Thus, we have:
$$
    \mathop{pcheck_{mv}}(\mathit{MVM},\phi) 
    \subseteq
    \biguplus_{i \in \{ 1, \dots, n\}}
     \{(i, m)| m\in \mathop{pcheck}(\mathop{proj}(\mathit{MVM},i),\phi)\}
$$

Any match $m: Q \rightarrow M_i$, where $M_i = \mathop{proj}(M_{\{1,\dots,n\}},i)$ is also structurally present in $\biguplus_{M_i \in M_{\{1, \dots, n\}}} M_i$. Since $\mathop{trans}_{mv}$ creates isomorphic graphs for $Q$ and the image of $Q$ under $m$, there exists a match $m_{mv} = \mathop{trans_{mv}}^m(m_{mv})$ for $Q_{mv} = \mathop{trans}_{mv}(Q)$ into $\mathit{MVM}$.

For $m$ to be reported by $\mathop{pcheck}(\mathop{proj}(M_{\{1,\dots,n\}},i),\phi)$, all elements in the image of $m$ have to be present in $M_i$. Hence, by the semantics of $\mathop{proj}$, we have that $\forall x \in Q: M_i \in p(\mathit{origin}_M^{-1}(m(x)))$ and hence, $\forall v \in V^{Q_{mv}}: M_i \in p(m_{mv}(v))$. Therefore, $\bigcap_{v \in V^{Q_{mv}}} p(m_{mv}(v)) \neq \emptyset$, so $(i, m) \in \mathop{pcheck_{mv}}(M_{\{1,\dots,n\}},\phi)$.

Thus, we have:
$$
  \mathop{pcheck_{mv}}(\mathit{MVM},\phi) 
  \supseteq 
  \biguplus_{i \in \{ 1, \dots, n\}}
  \{(i, m)| m\in \mathop{pcheck}(\mathop{proj}(\mathit{MVM},i),\phi)\}
$$

From the two subset-relations follows the equality. \qed
\end{proof}
\end{theorem}

\begin{theorem}

For a model versioning $\Delta^{M_{\{1, \dots, n\}}}$ and the associated multi-version model $\mathit{MVM} = \mathop{comb}(\Delta^M_{\{1, \dots, n\}})$ holds:

\begin{equation*}
\mathop{mcheck_{mv}}(\mathit{MVM}) =\biguplus_{(i,j,c) \in Y} \{(i, j, c, m) | m \in \mathop{mcheck}(m_{c, i}, m_{c, j})\},
\end{equation*}
where $Y = \{(i, j, c)\,|\,i, j \in \{ 1, \dots, n\}:  i < j, c \in \{c | M_c \in \mathop{pre^C}(i, j)\}\}$ and with $m_{c, i} = \mathop{proj^\Delta}(\mathit{MVM},c, i)$ and $m_{c, j} = \mathop{proj^\Delta}(\mathit{MVM},c, j)$.

\begin{proof}
For $e$ an edge and its invariant source $s$ reported for versions $M_i$ and $M_j$ and common predecessor $M_c \in \mathit{pre}^C(i, j)$ by $\mathop{mcheck}_{mv}$, by the normalization and the fact that $\mathit{pre}^C \neq \emptyset \rightarrow i \neq j$, we have that $i < j$. Because $M_c \notin p(v_e)$, we also have that $e \notin E^{M_c}$. Furthermore, we either have that $M_i \in p(\mathit{origin}^{-1}(e))$ and therefore $e \in E^{M_i}$ by Theorem \ref{the:proj} (case 1) or $M_j \in p(\mathit{origin}^{-1}(e))$ and therefore $e \in E^{M_j}$ (case 2). Since, $M_c \in \mathit{pre}(M_i)$, we therefore have a sequence of model modifications $M_c \Rightarrow^* M_i$ (case 1) or $M_c \Rightarrow^* M_j$ (case 2) that creates $e$.

We also know that, in case 1, the node in the multi-version model corresponding to $M_j$ lies on a path corresponding to a sequence of model modifications that deletes but does not recreate $s$. Hence, $M_j \notin p(\mathit{origin}^{-1}(s))$, since $M_j$ being in $p(\mathit{origin}^{-1}(s))$ would contradict the consistency of the model versioning. Furthermore, we know that $M_c \in p(\mathit{origin}^{-1}(s))$. This means there exists a sequence of model modifications $M_c \Rightarrow^* M_j$ that deletes $s$, but not $e$, since $e \notin E^{M_c}$. In case 2, by the same logic, there exists a sequence of model modifications $M_c \Rightarrow^* M_i$ that deletes $s$ but not $e$. Thus, there is an insert-delete conflict between $e$ and $s$ for $M_i$ and $M_j$ in either case.

The same is true for the target of an edge.

Thus, we have:
\begin{equation*}
\mathop{mcheck_{mv}}(\mathit{MVM}) \subseteq \biguplus_{(i,j,c) \in Y} \{(i, j, c, m) | m \in \mathop{mcheck}(m_{c, i}, m_{c, j})\},
\end{equation*}

If there exists an insert-delete conflict for an edge $e$ and its source $s$ between two versions $M_i = \mathop{proj}(M_{\{1,\dots,n\}},i)$ and $M_j = \mathop{proj}(M_{\{1,\dots,n\}},j)$ for a common predecessor $M_c \in \mathit{pre}^C(i, j)$, this means that there exist a sequence of model modifications $S_i = M_c \Rightarrow^* M_i$ and at least one sequence $S_j = M_c \Rightarrow^* M_j$, where one sequence deletes $s$ and does not delete $e$ and the other sequence creates $e$. Therefore, $s \in V^{M_c}$ and $e \notin E^{M_c}$.

In the case where $S_i$ is the sequence that creates $e$, it follows that $e \in E^{M_i}$. By Theoreme \ref{the:proj}, it then follows that $M_i \in p(v_e)$ and $M_c \notin p(v_e)$, with $v_e = \mathit{origin}^{-1}(e)$. Because $S_j$ then deletes and does not recreate $s$, it follows that there exists a path from a node connected to $v_s$ via a $dv$ edge to the node corresponding to $M_j$ that does not go through a node connected to $v_s$ via a $cv$ edge, with $v_s = \mathit{origin}^{-1}(s)$.

By the same logic, we can derive mirrored statements for the case where $S_j$ is the sequence that creates $e$.

From the existence of a sequence of model modifications that creates $e$, it also follows that $v_e$ is connected to some node $m_x$ via a $cv$ edge that does not correspond to $M_\alpha$. From the fact that one sequence deletes $e$ but not $s$, it follows that $p(v_e) \neq p(v_s)$.

Thus, $e$ and $v$ would be reported as an insert-delete conflict for versions $M_i$ and $M_j$ with predecessor $M_c$ in its normalized format by $\mathop{mcheck}_{mv}$.

The same is true for the target of an edge.

Thus, we have:
\begin{equation*}
\mathop{mcheck_{mv}}(\mathit{MVM}) \supseteq \biguplus_{(i,j,c) \in Y} \{(i, j, c, m) | m \in \mathop{mcheck}(m_{c, i}, m_{c, j})\},
\end{equation*}

By the two inclusions, we also have the equality. \qed
\end{proof}
\end{theorem}

\begin{theorem}
For a well-formedness constraint $\phi$, a correct model versioning $\Delta^{M_{\{1, \dots, n\}}}$ and the associated multi-version model $\mathit{MVM} = \mathop{comb}(\Delta^M_{\{1, \dots, n\}})$ holds:
\begin{equation*}
\mathop{pcheck^m_{mv}}(\mathit{MVM},\phi) = \biguplus_{(i, j, c) \in Y} \{(i, j, c, m) | m \in \mathop{pcheck}(M^{min}_{i, j, c}, \phi)\},
\end{equation*}
where $Y = \{(i, j, c)\,|\,i, j \in \{ 1, \dots, n\}:  i < j, c \in \{c | M_c \in \mathop{pre^C}(i, j)\}\}$ and $M^{min}_{i, j, c} = \mathop{merge_G^{min}}(\mathop{proj^\Delta}(\mathit{MVM},c, i),\mathop{proj^\Delta}(\mathit{MVM},c, j))$.

\begin{proof}
For a match $m$ for $Q$ into $\bigcup_{M_i \in M_{\{1,\dots,n\}}} M_i$ to be reported as a violation in the merge of versions $M_i$ and $M_j$ with latest common predecessor $M_c$ by $\mathop{pcheck^m_{mv}}$, there has to exist a corresponding match $m_{mv}: \mathop{trans}_{mv}(Q) \rightarrow \mathit{MVM}$. We know that, because $M_c \in \mathit{pre}^C(M_i, M_j)$, there are sequences of model modifications $S_i = M_c \Rightarrow^* M_i$ and $S_j = M_c \Rightarrow^* M_j$. Furthermore $m_{mv}$, we know that $\forall v \in m_{mv}(Q_{mv}): M_i \in p(v) \vee M_j \in p(v)$ and $\forall v \in m_{mv}(Q_{mv}): M_c \in p(v) \rightarrow (M_i \in p(v) \wedge M_j \in p(v))$.

Hence, every element $x \in m(Q)$ is either present in $M_c$ or created by $S_i$ or $S_j$, and none of these elements are deleted by either modification sequence. The resolution of merge conflicts by $\mathop{merge^{min}}$ may delete an edge created by one modification sequence, if its source or target is deleted by the other modification sequence. However, since $Q$ is a proper graph, meaning that for any edge $e \in E^Q: s^Q(e) \in V^Q \wedge t^Q(e) \in V^Q$, and no node $v \in m(V^Q)$ is deleted by either modification sequence, it follows that no edge $e \in m(E^Q)$ can be deleted in the process of resolving merge conflicts.

Hence, all elements $x \in m(Q)$ are present in $\mathop{merge_G^{min}}(S_i, S_j)$. By the construction of $MVM$, we also know that source and target functions of edges are preserved by $\mathop{trans}_{mv}^m(m_ {mv})$. This makes $m$ a violation of $\phi$ in the merge. Therefore, $(i, j, c, m) \in \mathop{pcheck}(\mathop{merge_G^{min}}(\mathop{proj}(M_{\{1,\dots,n\}},i),\mathop{proj}(M_{\{1,\dots,n\}},j)),\phi)$.

Since we know that $i < j$, we have:
\begin{equation*}
\mathop{pcheck^m_{mv}}(\mathit{MVM},\phi) \subseteq \biguplus_{(i, j, c) \in Y} \{(i, j, c, m) | m \in \mathop{pcheck}(M^{min}_{i, j, c}, \phi)\},
\end{equation*}

For two versions $M_i$ and $M_j$ with common predecessor $M_c = \mathit{pre}^C(M_i, M_j)$, a match $m \in \mathop{pcheck}(\mathop{merge_G^{min}}(\mathop{proj}(M_{\{1,\dots,n\}},i),\mathop{proj}(M_{\{1,\dots,n\}},j)),\phi)$ is also a valid match into $\bigcup_{M_i \in M_{\{1,\dots,n\}}} M_i$. There hence exists a corresponding match $m_{mv} : Q_{mv} \rightarrow \mathit{MVM}$ with $m = \mathop{trans}_{mv}^m(m_{mv})$.

Furthermore, in order to be present in the merge result, each element in the image of $m$ needs to be contained in $M_i$ or $M_j$ and, therefore, $\forall x \in m(\phi): x \in M_i \cup M_j$. We also know that none of the elements are deleted by the modification sequence $M_c \Rightarrow^* M_i$ or $M_c \Rightarrow^* M_i$, which means that $\forall x \in m(Q): x \in M_c \rightarrow (x \in M_i \wedge x \in M_j)$. We thus also have that $\forall v \in m_{mv}(V^{Q_{mv}}): M_i \in p(v) \vee M_j \in p(v)$ and $\forall v \in m_{mv}(V^{Q_{mv}}): M_c \in p(v) \rightarrow (M_i \in p(v) \wedge M_j \in p(v))$. Thereby, it must also hold that $M_i \in \argmin_{P \in \{ p(v) | v \in m_{mv}(V^{Q_{mv}}\}} |P| \vee M_j \in \argmin_{P \in \{ p(v) | v \in m_{mv}(V^{Q_{mv}}\}} |P|$. Because $i < j$, it follows that $(i, j, c, m) \in \mathop{pcheck^m_{mv}}(M_{\{1,\dots,n\}},\phi)$.

Thus, we have:
\begin{equation*}
\mathop{pcheck^m_{mv}}(\mathit{MVM},\phi) \supseteq \biguplus_{(i, j, c) \in Y} \{(i, j, c, m) | m \in \mathop{pcheck}(M^{min}_{i, j, c}, \phi)\},
\end{equation*}

By the two inclusions, we also have the equality. \qed
\end{proof}
\end{theorem}

\subsubsection{Complexity of $\mathop{pcheck_{mv}}$}

Effectively, the graphs $P_{mv}^M = \mathop{trans}_{mv}(P^M)$ and $Q_{mv} = \mathop{trans}_{mv}(Q)$ are only a different encoding of $P^M = \bigcup_{M_i \in M_{\{1, \dots, n\}}} M_i$ and $Q$, where edges are replaced by nodes with a single incoming and outgoing edge. Since these newly introduced nodes only cause a constant-factor overhead in pattern matching using an appropriate matching strategy and the size of the two graphs increases only by a constant factor, the effort for finding all matches for $Q_{mv}$ into $P_{mv}^M$ is in $O(C(\bigcup_{M_i \in M_{\{1, \dots, n\}}} M_i, Q))$.

The set $p(v_x)$ for each individual element $x$ with $v_x = \mathit{origin}^{-1}(x)$ can be computed via breadth-first searches starting in each node connected to $v_x$ via a $cv$ edge over $P^V_{mv}$ and stopping at nodes that are either connected to $v_x$ via a $dv$ edge or that have already been visited by a previous search. Since this ensures that the breadth-first searches do not perform redundant searches, the combined effort is in $O(|\Delta^{M_{1, \dots, n}}|)$. Given appropriate data structures, the effort for a single intersection of two version sets is in $O(|M_{\{1, \dots, n\}}|)$. Hence, computing $P = \bigcap_{v \in V^{Q_{mv}}} p(m_{mv}(v))$ for a single match $m_{mv}$ takes effort in $O(|V^{Q_{mv}}| \cdot |\Delta^{M_{1, \dots, n}}|)$.

For a number of matches $X$ for $Q_{mv}$ into $P_{mv}^M$, the effort for $\mathop{pcheck_{mv}}$ is then in $O(C(\bigcup_{M_i \in M_{\{1, \dots, n\}}} M_i, Q) + X \cdot |V^{Q_{mv}}| \cdot |\Delta^{M_{1, \dots, n}}|)$.

\subsubsection{Complexity of $\mathop{mcheck_{mv}}$}

We can precompute the latest common predecessors for any pair of versions $M_i$ and $M_j$ according to the definition in $O(|P^V_{mv} \cdot |M_{\{1,\dots,n\}}| + |M_{\{1,\dots,n\}}|^4)$ via breadth-first search for finding all predecessors for every version and basic set operations. Since the maximally preserving model modification between two versions along identity morphisms is uniquely defined and therefore $|E^{P^V_{mv}}| \leq |M_{\{1,\dots,n\}}|^2$, the complexity is in $O(|M_{\{1,\dots,n\}}|^4)$. This computation allows an efficient enumeration of the set of model modification pairs representing possible merges.

Since information about creation and deletion of elements is not explicitly available in a model-modification-based representation of a versioning, finding all insert-delete conflicts between two model modifications via $\mathop{mcheck}$ can be done by checking for each edge in either modification's resulting model whether it is created by that modification and its source or target is deleted by the other modification. Given appropriate data structures for storing the individual model versions, the check for one edge can be performed in constant time thanks to the requirement of partial identity morphisms. This yields a computational complexity in $O(|M_{\{1,\dots,n\}}|^4 + \sum_{(\gspan{M_c}{K_i}{M_i}, \gspan{M_c}{K_j}{M_j}) \in Y} |E^{M_i}| + |E^{M_j}|)$ for checking an entire model versioning with $Y$ the set of possible merges represented by pairs of model modifications. Note that, in the worst case, there may exist up to $O(|M_{\{1,\dots,n\}}|^3)$ possible merges in a model versioning, so in the worst case, this implies effort in $O(|M_{\{1,\dots,n\}}|^4 + |E^{M_{max}}| \cdot |M_{\{1,\dots,n\}}|^3)$, where $|E^{M_{max}}|$ is the maximum number of edges present in a single model version.

The computation of $P = p(v_e)$ and $D$ for every node $v_e \in V^{MVM}$, where $\mathit{corr}_{mv}^{-1}(type^{MVM}(v_e)) \in E^{TM}$ and $v_e$ has a $cv$ edge to a node representing another version than $M_\alpha$, and the edge source $v_s = s^M(\mathit{origin}(v_e))$ can be performed in $O(\Delta_+ \cdot |P^V_{mv}|)$, where $P^V_{mv}$ is the version graph and $\Delta_+$ is the overall number of elements created by the model versioning. Checking whether $p(v_e) = p(v_s)$ only takes effort in $O(|M_{\{1,\dots,n\}}|)$.

For a node $v_e$ and versions $M_i \in P$ and $M_j \in D$, we can then determine all latest predecessors for which $M_i$ and $M_j$ are in conflict in $O(|M_{\{1,\dots,n\}}|)$. Since there are at most $|M_{\{1,\dots,n\}}|^2$ pairs of versions in $P \times D$, the overall complexity of $\mathop{mcheck}_{mv}$ considering the effort for precomputing latest common predecessors is in $O(|M_{\{1,\dots,n\}}|^4 + \Delta_+ \cdot |M_{\{1,\dots,n\}}|^3)$, which simplifies to $O(|\Delta_+ \cdot |M_{\{1,\dots,n\}}|^3)$ if $|M_{\{1,\dots,n\}}| \leq \Delta_+$. The computation for edge targets works analogously and thus does not change the complexity.

\subsubsection{Complexity of $\mathop{pcheck^m_{mv}}$}

We denote the effort for finding all matches for a pattern $Q$ in a model $M_m$ by $C(M_m, Q)$. A possible merge in a model versioning is characterized by two model modifications, that is, two model versions $M_i$ and $M_j$ and a latest common predecessor $M_c$. Disregarding the effort for computing a merge itself, the effort for finding all violations of a well-formedness condition characterized by $Q$ in a model versioning with $Y$ the set of possible merges via computing all merges and searching for $Q$ in the results is then given by $O(\sum_{(m_1, m_2) \in Y} C(\mathop{merge^{min}_G}(m_1, m_2), Q))$. Potentially, there are up to $O(|M_{\{1, \dots, n\}}|^3)$ possible merges in a model versioning with versions $M_{\{1, \dots, n\}}$, which can be computed in $O(|M_{\{1,\dots,n\}}|^4)$. This results in a complexity for computing and checking all these merges via $\mathop{pcheck}$ in $O(|M_{\{1,\dots,n\}}|^4 + \sum_{(m_1, m_2) \in Y} C(\mathop{merge^{min}_G}(m_1, m_2), Q))$.

Finding all violations characterized by a pattern $Q$ in a model versioning by $\mathop{pcheck^m_{mv}}$ requires a search for $Q_{mv}$ over the corresponding multi-version model. As discussed in Section \ref{sec:pcheck_mv_complexity}, this requires effort in $O(C(\bigcup_{M_i \in M_{\{1, \dots, n\}}} M_i, Q))$. Then, for each structural match that is found, the set $P_{v} = p(v)$ has to be computed for each $v \in V^{Q_{mv}}$, which takes $O(|V^{Q_{mv}}| \cdot |P^V_{mv}|)$ computational steps. Finding all complementary versions $M_j$ regarding $Q_{mv}$ for a version $M_i$ requires $O(|V^{Q_{mv}}| \cdot |M_{\{1, \dots, n\}}|)$ steps.

The number of candidates of complementary versions is in $O(|M_{\{1,\dots,n\}}|^2)$. Latest common predecessors for all version pairs can be computed and cached upfront in $O(|M_{\{1,\dots,n\}}|^4)$. For each latest common predecessor, the remaining checks can be performed in $O(|V^{Q_{mv}}|)$ given appropriate data structures. Since any pair of versions has at most $|M_{\{1, \dots, n\}}|$ common predecessors, we thus have effort in $O(|V^{Q_{mv}}| \cdot |M_{\{1, \dots, n\}}|^3)$ for each match.

For $X$ matches for $Q_{mv}$ into $\mathit{MVM}$, the overall computational effort is thus in $O(|M_{\{1,\dots,n\}}|^4 + C(\bigcup_{M_i \in M_{\{1, \dots, n\}}} M_i, Q) + X \cdot |V^{Q_{mv}}| \cdot |M_{\{1, \dots, n\}}|^3)$.

\newpage


\section{Violation Patterns of Considered Well-Formedness Constraints}

\begin{figure}
\centering
\includegraphics[width=0.65\textwidth]{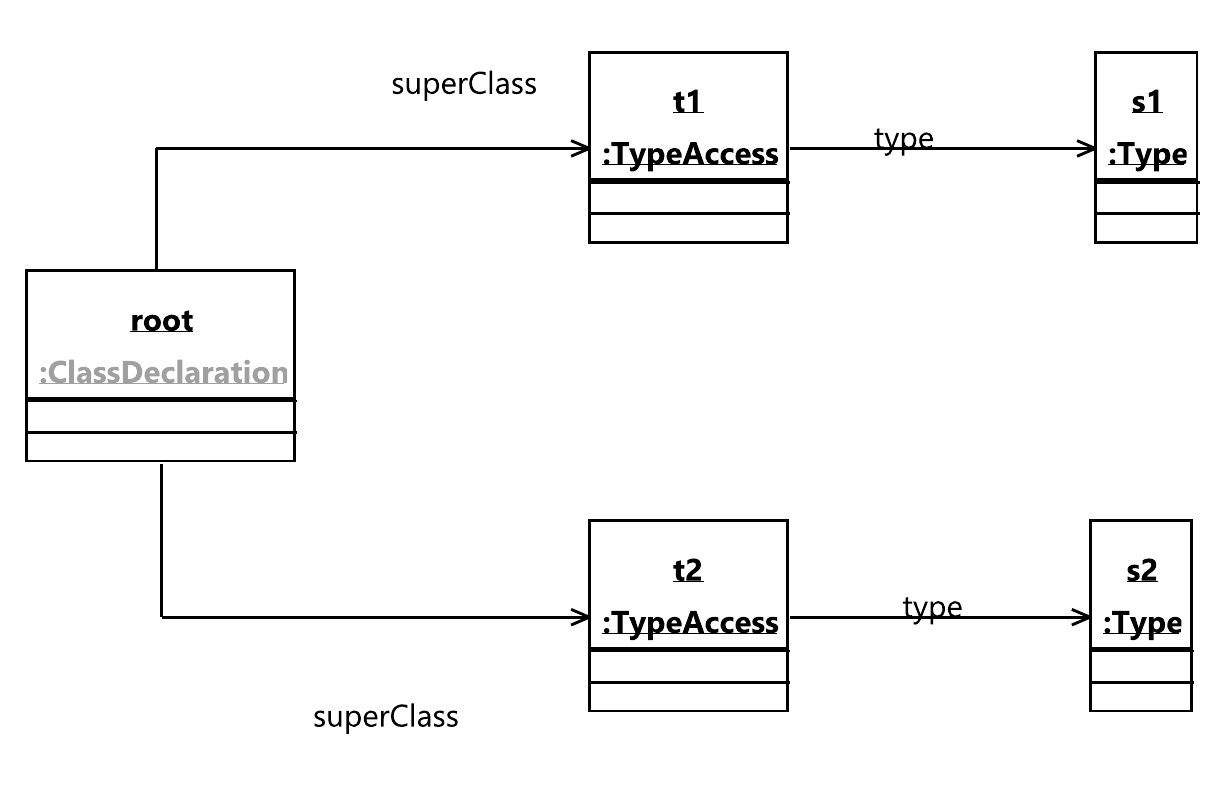}
\caption{Violation pattern for the constraint ``unique superclass''}
\end{figure}

\begin{figure}
\centering
\includegraphics[width=0.65\textwidth]{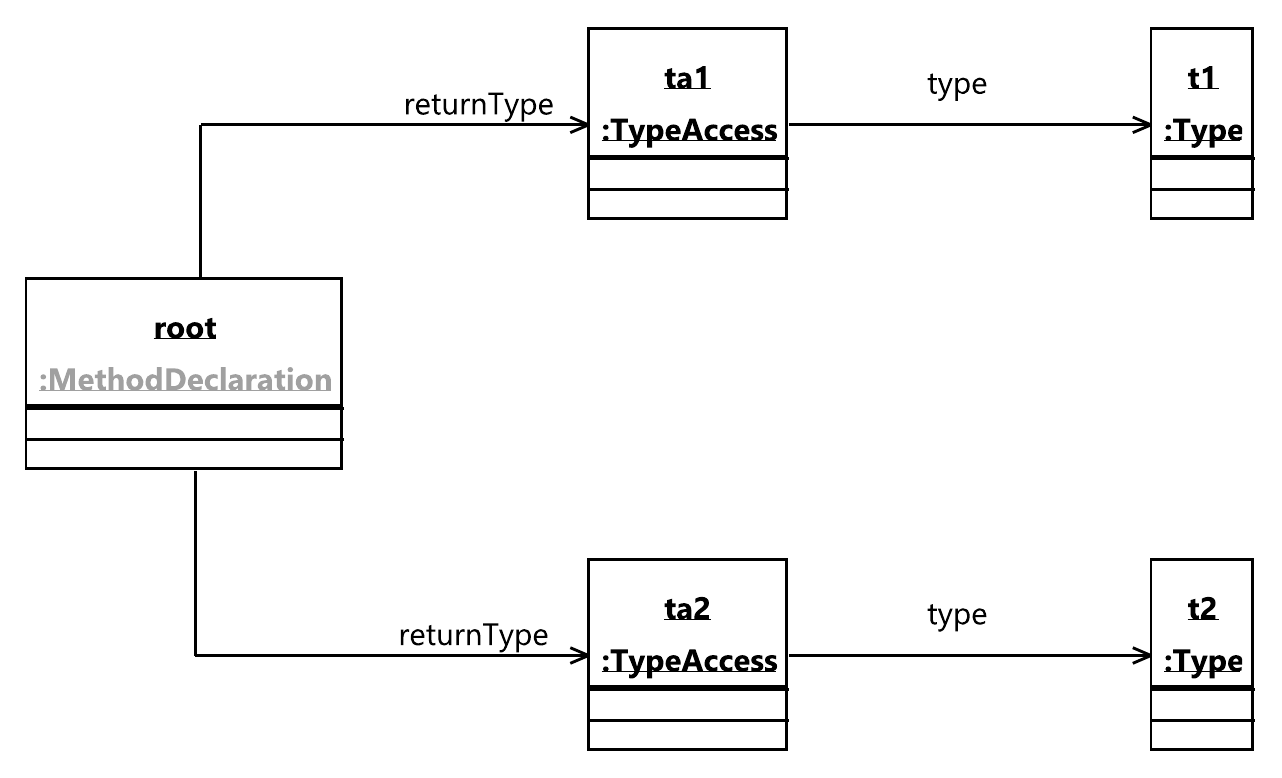}
\caption{Violation pattern for the constraint ``unique return type''}
\end{figure}

\begin{figure}
\centering
\includegraphics[width=0.65\textwidth]{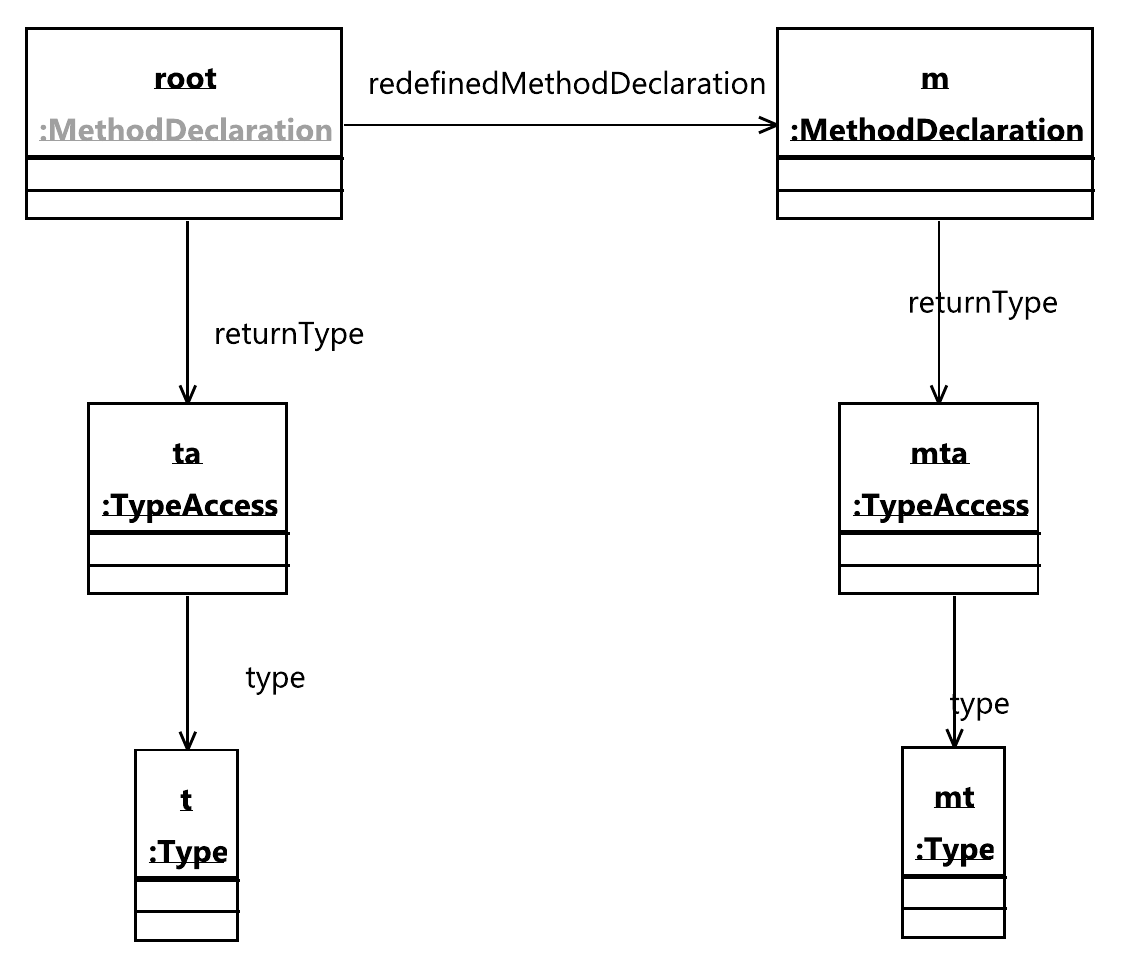}
\caption{Violation pattern for the constraint ``consistent override''}
\end{figure}

\end{document}